\documentclass[letterpaper,twocolumn,10pt]{article}
\usepackage{usenix2019_v3}

\usepackage{amsmath}
\usepackage{booktabs}
\usepackage{multirow}
\usepackage{array}
\usepackage{tabularx}
\usepackage{graphicx}
\usepackage{xurl}  
\usepackage{tcolorbox}  
\usepackage{listings}   
\usepackage{float}      
\usepackage{threeparttable}
\usepackage{pifont}  
\usepackage{xcolor}  
\usepackage{tikz}    
\usetikzlibrary{tikzmark,arrows.meta}
\newcommand{\cmark}{{\color{green!70!black}\ding{51}}}  
\newcommand{\xmark}{{\color{red}\ding{55}}}  
\begin{document}

\date{}

\title{\Large \bf FuzzingBrain V2: A Multi-Agent LLM System for Automated Vulnerability Discovery and Reproduction}

\author{
{\rm Ze Sheng}\\
Texas A\&M University\\
zesheng@tamu.edu
\and
{\rm Zhicheng Chen}\\
Texas A\&M University\\
chenzc2001@tamu.edu
\and
{\rm Qingxiao Xu}\\
Texas A\&M University\\
qingxiao@tamu.edu
\and
{\rm Kewen Zhu}\\
Texas A\&M University\\
steve1998kewen@gmail.com
\and
{\rm Jeff Huang}\\
Texas A\&M University\\
jeffhuang@tamu.edu
}

\maketitle

\begin{abstract}
Software vulnerabilities pose critical security threats, with nearly 50,000 CVEs reported in 2025. While Large Language Models (LLMs) show promise for automated vulnerability detection, three key challenges remain. First, LLM-generated vulnerability reports suffer from high false positive rates and lack reproducible verification. Second, existing LLM-based approaches use suboptimal granularities for vulnerability localization: function-level analysis overlooks bugs when context becomes extensive, while line-level analysis lacks sufficient context. Third, existing approaches have difficulty reasoning about vulnerabilities with complex cross-function dependencies and triggering conditions.

We present FuzzingBrain V2, a multi-agent system that addresses these gaps through four key contributions: (1) fully automated vulnerability analysis built on Google's OSS-Fuzz, ensuring all reported vulnerabilities are fuzzer-reproducible; (2) Suspicious Point, a novel control-flow-based abstraction for precise vulnerability localization at the optimal granularity; (3) logic-driven hierarchical function analysis with dual-layer fuzzing enhancing function coverage under resource constraints; (4) MCP-based static and dynamic analysis tools with context engineering enhancing complex vulnerability reasoning.

On the AIxCC 2025 Final Competition C/C++ dataset, FuzzingBrain V2 achieved 90\% detection rate (36 of 40 vulnerabilities). In real-world deployment, FuzzingBrain V2 discovered 29 zero-day vulnerabilities across 12 open-source projects, all confirmed and fixed by maintainers, with 2 assigned CVE IDs.
\end{abstract}

\section{Introduction}

Software vulnerabilities remain a critical threat to system security. CVE disclosures have surged in recent years, with over 40,000 new vulnerabilities reported in 2024 and nearly 50,000 in 2025. Moreover, recent studies indicate that developers using AI coding assistants tend to produce less secure code while exhibiting overconfidence in its safety~\cite{perry2023,llmcyber2025}. Studies from Microsoft and Google indicate that approximately 70\% of security vulnerabilities in C/C++ codebases stem from memory safety issues~\cite{microsoft70,chromium70}, a finding reinforced by CISA's 2025 guidance on memory safe languages~\cite{cisa2025}. The MongoBleed vulnerability (CVE-2025-14847) in December 2025, which exposed over 87,000 MongoDB servers to unauthenticated memory leaks, underscores how memory safety issues continue to threaten critical infrastructure~\cite{mongobleed}. Effective vulnerability detection is therefore essential for modern software development.

\begin{figure}[t]
\centering
\includegraphics[width=0.9\columnwidth]{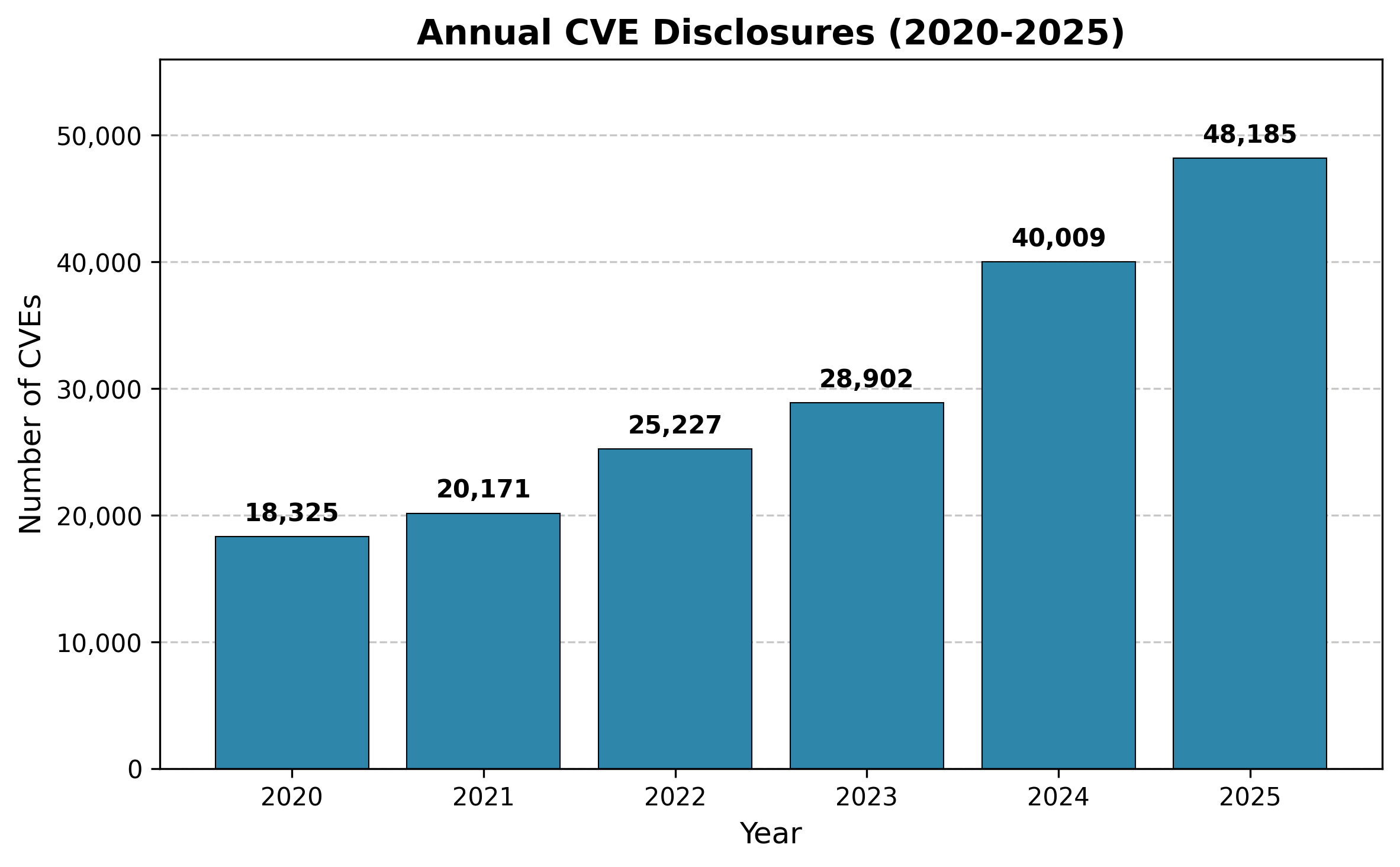}
\caption{Annual CVE disclosures from 2020 to 2025, showing a sharp increase in reported vulnerabilities.}
\label{fig:cve_trend}
\end{figure}

State-of-the-art vulnerability detection approaches fall into two categories: static analysis and dynamic analysis. Static analysis tools leverage techniques such as code property graphs~\cite{joern}, abstract interpretation~\cite{infer}, and query-based analysis~\cite{codeql} to examine source code for potential vulnerabilities. However, these approaches suffer from high false positive rates that overwhelm manual review. Dynamic analysis, particularly fuzz testing with tools like AFL~\cite{afl} and libFuzzer~\cite{libfuzzer}, discovers vulnerabilities through input mutation and crash monitoring. However, fuzzing operates without semantic understanding of program logic, limiting its ability to reach deep vulnerabilities protected by complex path constraints such as checksums, magic numbers, or multi-stage protocol handshakes.

Large Language Models (LLMs) offer a potential solution by combining code comprehension with reasoning capabilities. Models trained on code corpora can understand program semantics, identify suspicious patterns, and reason about data flow across functions. The success of DARPA's AIxCC 2025 competition~\cite{aixcc}, where LLM-based systems discovered real vulnerabilities in production software, demonstrates this potential. However, directly applying LLMs to vulnerability detection faces significant challenges.

According to recent surveys on LLM-based vulnerability detection~\cite{sheng2025llmsecurity,zhou2024llmvuln}, three fundamental gaps remain:

\smallskip\noindent\textbf{Gap 1:} LLM-reported vulnerabilities cannot be verified without executable proof-of-concept inputs, leaving security teams unable to distinguish true vulnerabilities from hallucinations.

\smallskip\noindent\textbf{Gap 2:} Current approaches operate at suboptimal granularities. Function-level analysis is too coarse—LLMs exhibit attention bias toward certain patterns while overlooking others. Line-level analysis is too fine—insufficient context leads to high false positive rates.

\smallskip\noindent\textbf{Gap 3:} Complex contextual dependencies—cross-function data flow, indirect calls, and stateful operations—limit LLM understanding of vulnerabilities spanning multiple functions.

We propose FuzzingBrain V2, an MCP-based multi-agent vulnerability detection system that addresses these gaps. For \textbf{Gap 1}, we adopt Google's OSS-Fuzz framework as our verification backend, ensuring that every reported vulnerability is reproducible through established fuzzers and sanitizers—OSS-Fuzz integrates over 1,000 open-source projects, enabling direct deployment at scale. For \textbf{Gap 2}, we introduce Suspicious Point (SP), a novel abstraction between line-level and function-level granularity that captures vulnerability-relevant code regions with control flow context, enabling both precise localization and systematic reproduction. For \textbf{Gap 3}, we employ a multi-agent architecture with hierarchical search strategy, where specialized agents collaborate to gather cross-function context, analyze complex data flows, and reason about program state. Our system leverages LLM code comprehension to analyze business logic and identify suspicious behaviors that deviate from expected program semantics.

This paper makes the following contributions:
\begin{itemize}
\item We present FuzzingBrain V2, a fully automated multi-agent vulnerability detection system built on Google's OSS-Fuzz and the Model Context Protocol (MCP). The system can be directly deployed on over 1,000 open-source projects, guarantees 100\% reproducibility for confirmed vulnerabilities, and automatically generates submission-ready vulnerability reports.

\item We introduce Suspicious Point (SP), a novel abstraction between line-level and function-level granularity that captures vulnerability-relevant code regions with control flow context, bridging the gap between LLM analysis and fuzzing-based reproduction.

\item We propose a logic-driven hierarchical search strategy with dual-layer fuzzing. Rather than matching known vulnerability patterns, our system analyzes business logic to identify suspicious behaviors, enabling deep understanding of vulnerabilities with complex contextual dependencies.

\item We evaluate FuzzingBrain V2 on the AIxCC C/C++ dataset, achieving 90\% detection rate (36 of 40 vulnerabilities), and discovered 41 zero-day vulnerabilities across 19 open-source projects.
\end{itemize}

The remainder of this paper is organized as follows: Section~\ref{sec:background} provides background on fuzzing and LLM agents. Section~\ref{sec:design} describes FuzzingBrain V2's system design. Section~\ref{sec:impl} discusses implementation details. Section~\ref{sec:eval} presents our evaluation. Section~\ref{sec:discussion} discusses limitations and future work. Section~\ref{sec:conclusion} concludes.

\section{Background and Related Work}
\label{sec:background}

\subsection{Fuzzing and OSS-Fuzz}

Fuzz testing (fuzzing) is a dynamic analysis technique that discovers software vulnerabilities by feeding programs with randomly generated or mutated inputs and monitoring for abnormal behaviors. Modern fuzzers such as AFL~\cite{afl} and libFuzzer~\cite{libfuzzer} employ coverage-guided mutation strategies, prioritizing inputs that explore new code paths to maximize code coverage. To detect vulnerabilities beyond simple crashes, fuzzers are typically paired with sanitizers—compiler-based instrumentation tools that detect memory errors at runtime. AddressSanitizer (ASan) detects out-of-bounds accesses and use-after-free errors, MemorySanitizer (MSan) identifies uninitialized memory reads, and UndefinedBehaviorSanitizer (UBSan) catches undefined behavior such as integer overflows.

OSS-Fuzz~\cite{ossfuzz} is Google's continuous fuzzing infrastructure for open-source software. We adopt OSS-Fuzz as our fuzzing backend for three reasons: (1) \textbf{Easy integration}: OSS-Fuzz integrates over 1,000 open-source projects, allowing FuzzingBrain V2 to be directly deployed without additional configuration; (2) \textbf{Extensibility}: custom fuzzers and harnesses can be easily added to target specific functions or code paths; (3) \textbf{Built-in analysis toolchain}: OSS-Fuzz integrates Fuzz Introspector, which provides call graph and reachability analysis along with runtime coverage information, which we leverage for code understanding and coverage-guided exploration. In this work, we focus on C/C++ projects and use libFuzzer as the primary fuzzing engine.

\subsection{Vulnerability Definition}

In this paper, we adopt a fuzzing-centric definition of vulnerability. Given a program $P$ and a code location $v$:
\begin{equation}
v \in \mathcal{V}(P) \iff \exists \, i : \forall \, \textit{run}, \; \textit{loc}(\textit{crash}(\textit{exec}(P, i))) = v
\end{equation}
A location $v$ is vulnerable if and only if there exists an input $i$ that consistently triggers a sanitizer-detected crash at $v$ across all executions. This definition ensures every reported vulnerability is verifiable and reproducible, eliminating the uncertainty inherent in static analysis or LLM-only approaches.

\subsection{LLM Agents and Model Context Protocol}

An LLM agent is an autonomous system that combines a large language model with external tools and iterative reasoning to accomplish complex tasks. Unlike simple LLM queries, agents can plan multi-step actions, invoke tools (e.g., code search, compilers, fuzzers), observe results, and adapt their strategy accordingly. Multi-agent systems extend this concept by coordinating multiple specialized agents, each responsible for a specific subtask, to solve problems beyond the capability of a single agent.

The Model Context Protocol (MCP)~\cite{mcp} is an open standard by Anthropic for LLM-tool integration. Without MCP, each AI application requires custom code to connect to each tool. MCP defines a unified interface: AI applications implement a Client, tools implement a Server, and they interoperate automatically.

We adopt FastMCP~\cite{fastmcp} as the MCP framework for FuzzingBrain V2: (1) FuzzingBrain V2 itself is wrapped as an MCP-callable tool, allowing external LLM applications to invoke vulnerability detection as a service; (2) all internal agents follow the MCP protocol, enabling modular design and seamless inter-agent communication.

\subsection{Related Work}

\textbf{LLM-based Vulnerability Detection.}
Recent surveys~\cite{sheng2025llmsecurity,zhou2024llmvuln} provide comprehensive overviews of this rapidly evolving field. Early work applied deep learning to vulnerability detection: VulDeePecker~\cite{vuldeepecker} pioneered using BLSTM on code slices, while Devign~\cite{devign} leveraged graph neural networks to learn from AST, CFG, and DFG structures. Pre-trained language models further advanced this field: CodeBERT~\cite{codebert} introduced bimodal pre-training on code and natural language, and VulBERTa~\cite{vulberta} specialized RoBERTa for C/C++ vulnerability detection. Recent work has examined LLM capabilities more critically—Ding et al.~\cite{ding2024} systematically evaluated code LLMs and found their vulnerability detection performance falls short of expectations. GRACE~\cite{grace} attempts to improve LLM detection by combining graph structures with in-context learning. Li et al.~\cite{li2025correct} demonstrated that context is crucial for LLM-based detection, achieving significant improvements when sufficient context is provided. Huynh et al.~\cite{huynh2025dsn} explored prompt engineering strategies for vulnerability detection. LLMxCPG~\cite{llmxcpg} uses code property graphs to extract minimal context slices for LLM analysis. However, these approaches primarily treat vulnerability detection as a classification task without providing reproducible proof-of-concept inputs.

\textbf{LLM-based Fuzzing.}
Recent work has explored using LLMs to enhance fuzzing. Fuzz4All~\cite{fuzz4all} uses LLMs to generate test inputs across multiple languages. WhiteFox~\cite{whitefox} applies LLMs to white-box compiler fuzzing, generating inputs that trigger optimization bugs. KernelGPT~\cite{kernelgpt} uses LLMs to generate syscall specifications for Syzkaller. Most relevant to our work, OSS-Fuzz-Gen~\cite{ossfuzzgen} uses LLMs to automatically generate fuzz drivers for OSS-Fuzz. While these approaches improve fuzzing efficiency, they focus on input or harness generation rather than vulnerability-directed exploration.

\textbf{Hybrid Approaches.}
To improve reliability, recent work combines LLMs with traditional analysis tools. IRIS~\cite{iris} uses LLMs to filter false positives from static analysis. LLM4Vuln~\cite{llm4vuln} provides a unified framework to evaluate and enhance LLM vulnerability reasoning. Vul-RAG~\cite{vulrag} uses retrieval-augmented generation to incorporate known vulnerability knowledge. LLM-SmartAudit~\cite{llmsmartaudit} introduces a multi-agent architecture for smart contract vulnerability detection. Unlike these approaches, FuzzingBrain V2 integrates LLM analysis with fuzzing-based verification for C/C++ programs, ensuring every reported vulnerability is reproducible through crash-triggering inputs.

\section{System Design}
\label{sec:design}

\subsection{Overview}

\begin{figure*}[t]
\centering
\includegraphics[width=\textwidth]{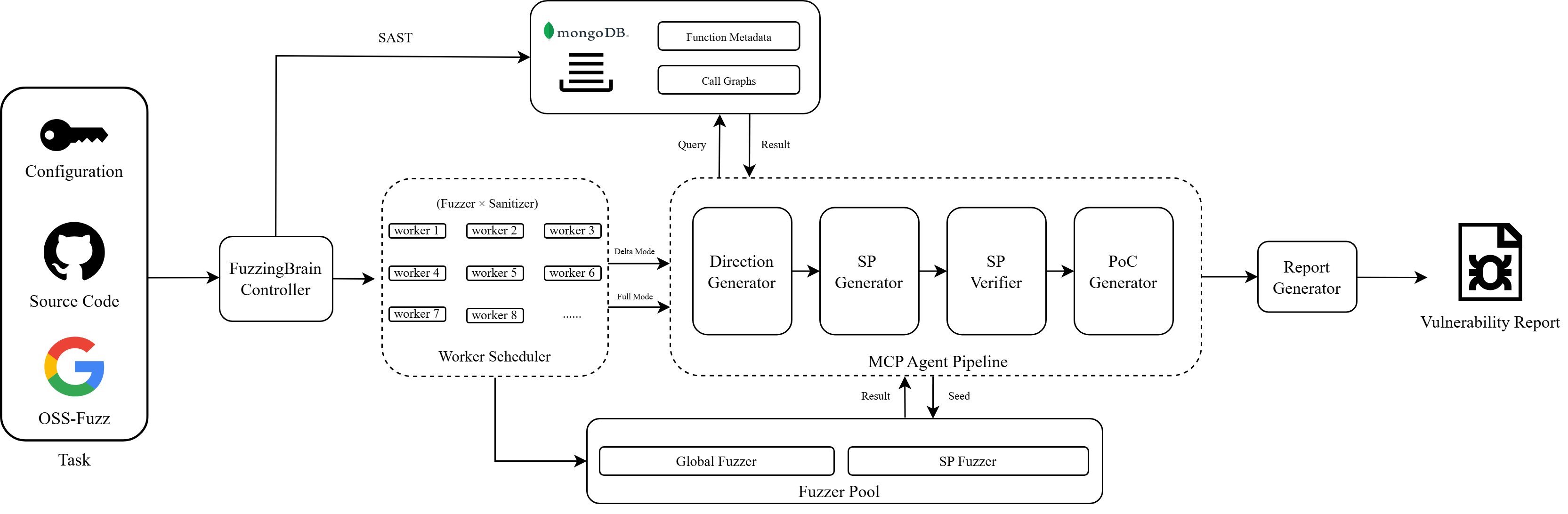}
\caption{System overview of FuzzingBrain V2. The Controller orchestrates static analysis, agent pipeline execution, and fuzzing. Components communicate through a shared database, with the SP abstraction bridging SP Generator, SP Verifier, and PoC Generator.}
\label{fig:overview}
\end{figure*}

Figure~\ref{fig:overview} presents the architecture of FuzzingBrain V2. The system operates in three stages: (1) \textit{Static Analysis} extracts function metadata and call graphs via Fuzz Introspector, storing results in a database; (2) \textit{Agent Pipeline} employs specialized LLM agents to discover and verify Suspicious Points through hierarchical search; (3) \textit{PoC Generation} combines PoC Generator reasoning with dual-layer fuzzing to generate reproducible crash-triggering inputs.

The system takes an OSS-Fuzz project as input and outputs vulnerability reports with crash-triggering inputs. Each worker handles a specific fuzzer-sanitizer pair (e.g., \texttt{libpng\_read\_fuzzer} with ASan), enabling parallel analysis across different sanitizers. FuzzingBrain V2 supports two scan modes:

\begin{itemize}
\item \textbf{Full-Scan Mode}: Given a fuzzer, performs comprehensive vulnerability analysis across the entire codebase.
\item \textbf{Delta-Scan Mode}: Given a fuzzer and two code versions (e.g., a commit), analyzes whether the changes introduce new vulnerabilities.
\end{itemize}

Table~\ref{tab:agents} lists the model tiers and specialized agents in FuzzingBrain V2.

\begin{table*}[t]
  \centering
  \caption{Model tiers and specialized agents in FuzzingBrain V2.}
  \label{tab:agents}
  \small
  \setlength{\tabcolsep}{8pt}
  \begin{tabular}{lll}
    \toprule
    \textbf{Tier} & \textbf{Models} & \textbf{Usage} \\
    \midrule
    T1 (Reasoning) & O3, GPT-5.2-Pro, Claude-Opus-4.5 & Complex reasoning, strategic planning \\
    T2 (Main) & Claude-Sonnet-4.5, GPT-5.2, Gemini-3-Pro & Code analysis, core agent work \\
    T3 (Utils) & Claude-Haiku-4.5, GPT-5-Mini, Gemini-3-Flash & Context compression \\
    \bottomrule
  \end{tabular}

  \vspace{8pt}

  \begin{tabular}{llll}
    \toprule
    \textbf{Agent} & \textbf{Role} & \textbf{Model Tier} & \textbf{Section} \\
    \midrule
    Direction Generator & Divide codebase into logical directions & T1 & \ref{sec:direction-def} \\
    SP Generator & Initial screening for suspicious patterns & T2 & \ref{sec:spgen} \\
    SP Deduplicator & Identify and merge duplicate SPs & T3 & \ref{sec:spgen} \\
    SP Verifier & In-depth verification of SPs & T1 & \ref{sec:spverify} \\
    PoC Generator & Craft inputs to trigger vulnerabilities & T2 & \ref{sec:pocgen} \\
    Report Agent & Refine SP and generate vulnerability report & T2 & \ref{sec:pocgen} \\
    Seed Generator & Generate fuzzer seeds from directions/FPs & T2 & \ref{sec:fuzzing} \\
    Context Compressor & Compress context for cross-function analysis & T3 & \ref{sec:spverify} \\
    \bottomrule
  \end{tabular}
\end{table*}

\subsection{Suspicious Point}

During logic-driven search (Section~\ref{sec:direction}), the SP Generator analyzes functions to identify potential vulnerabilities. Each identified vulnerability is captured as a \textit{Suspicious Point (SP)}—the core abstraction that bridges SP Generator, SP Verifier, and PoC Generator.

SP addresses the granularity problem in LLM-based vulnerability detection. Rather than asking the LLM to output a vague analysis of an entire function, SP requires the LLM to produce a structured report for each suspicious code location—treating it as a real vulnerability. Each SP is independent, enabling parallel verification and PoC generation.

\subsubsection{SP Structure and Lifecycle}

\begin{figure}[h!]
\begin{tcolorbox}[colback=blue!5, colframe=blue!40, title=Suspicious Point Example]
\small
\textbf{function}: \texttt{png\_read\_row} (\texttt{pngread.c})\\[2pt]
\textbf{sources}: [\texttt{png\_fuzzer} + ASan, \texttt{png\_read\_fuzzer} + ASan]\\[2pt]
\textbf{description}: In the \texttt{else} branch of the \texttt{if (png\_ptr->transformations)} check, after calling \texttt{png\_do\_read\_transformations()}, the \texttt{memcpy} at the end of the function copies \texttt{row\_bytes} into \texttt{row} without verifying that \texttt{row} has sufficient space when \texttt{row\_bytes} exceeds the originally allocated size.\\[2pt]
\textbf{vuln\_type}: heap-buffer-overflow\\[2pt]
\textbf{score}: 0.82\\[2pt]
\textbf{is\_verified}: true\\[2pt]
\textbf{poc\_guidance}: Craft PNG with large \texttt{row\_bytes} after transformation; set \texttt{transformations} flag to trigger the else branch.\\[2pt]
\textbf{is\_real}: true\\[2pt]
\textbf{poc\_attempted\_by}: [\texttt{png\_fuzzer}, \texttt{png\_read\_fuzzer}]\\[2pt]
\textbf{poc\_ids}: [\texttt{a3f2c1d8-...}, \texttt{b7e4f9a2-...}]
\end{tcolorbox}
\caption{Example Suspicious Point. The description uses control-flow landmarks instead of line numbers. Fields are populated progressively: \texttt{is\_verified} and \texttt{poc\_guidance} after verification; \texttt{is\_real}, \texttt{poc\_attempted\_by}, and \texttt{poc\_ids} after PoC generation.}
\label{fig:sp-example}
\end{figure}

Figure~\ref{fig:sp-example} shows an example SP. We use control-flow descriptions instead of line numbers because LLMs frequently hallucinate exact line numbers.

An SP progresses through three stages (see Figure~\ref{fig:overview}), with fields populated progressively:
\begin{itemize}
\item \textbf{Creation}: The SP Generator analyzes functions for suspicious patterns and performs preliminary feasibility checks, populating \texttt{function}, \texttt{sources}, \texttt{description}, \texttt{vuln\_type}, and initial \texttt{score}.
\item \textbf{Verification}: The SP Verifier performs deeper analysis, tracing call paths and checking for safety boundaries, then updates \texttt{is\_verified}, \texttt{score}, and \texttt{poc\_guidance}.
\item \textbf{PoC Generation}: The PoC Generator crafts inputs for fuzzer and sanitizer to trigger the vulnerability, setting \texttt{is\_real}, \texttt{poc\_attempted\_by}, and \texttt{poc\_ids} upon success.
\end{itemize}

\subsection{Task Processing and Worker Distribution}
\label{sec:task}

\subsubsection{Static Analysis and Call Graph}
\label{sec:static}

FuzzingBrain V2 begins by constructing a call graph from static analysis results. While we use Fuzz-Introspector~\cite{ossfuzz} by default, the system accepts any tool that provides function metadata and call relationships. The static analysis extracts: (1) function metadata (name, file path, source code), (2) call relationships (caller-callee edges), and (3) fuzzer reachability (which fuzzers can reach each function). From this data, we construct a global call graph containing all functions reachable by at least one fuzzer. Each function maintains a \texttt{reached\_by\_fuzzers} field indicating which fuzzers can reach it. We compute \texttt{call\_depth} for each function via BFS from entry points (functions directly called by fuzzers).

\subsubsection{Worker Distribution}
\label{sec:worker}

\begin{figure}[t]
\centering
\includegraphics[width=0.9\columnwidth]{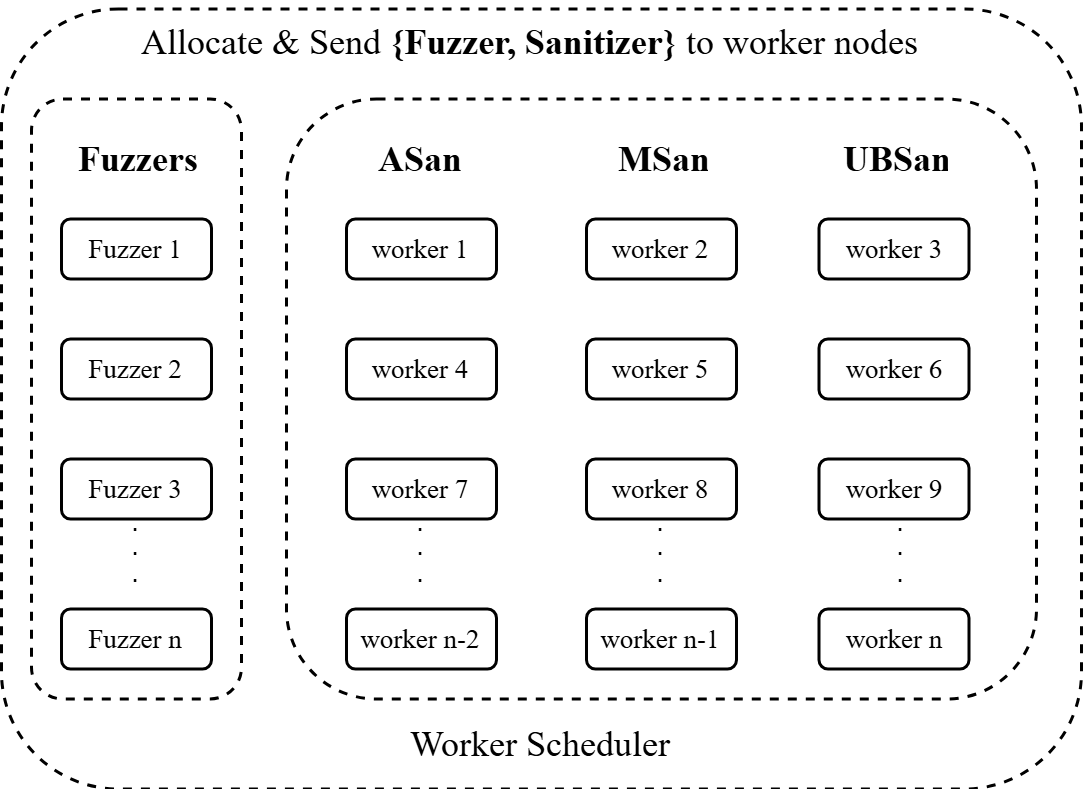}
\caption{Worker distribution. The scheduler allocates each (fuzzer, sanitizer) pair to a separate worker, enabling parallel analysis across different fuzzers and sanitizers.}
\label{fig:worker}
\end{figure}

After static analysis, FuzzingBrain V2 distributes work across parallel workers. Each worker handles a specific \textit{task}—a (fuzzer, sanitizer) pair such as (\texttt{png\_read\_fuzzer}, AddressSanitizer). This design reflects a key insight: different fuzzers reach different code paths, and different sanitizers detect different vulnerability classes. For a project with $F$ fuzzers and $S$ sanitizers, we spawn up to $F \times S$ workers (Figure~\ref{fig:worker}). Each worker operates on a filtered subgraph of the global call graph containing only functions reachable by its assigned fuzzer, then executes the full agent pipeline independently.

\begin{figure*}[t]
\centering
\includegraphics[width=\textwidth]{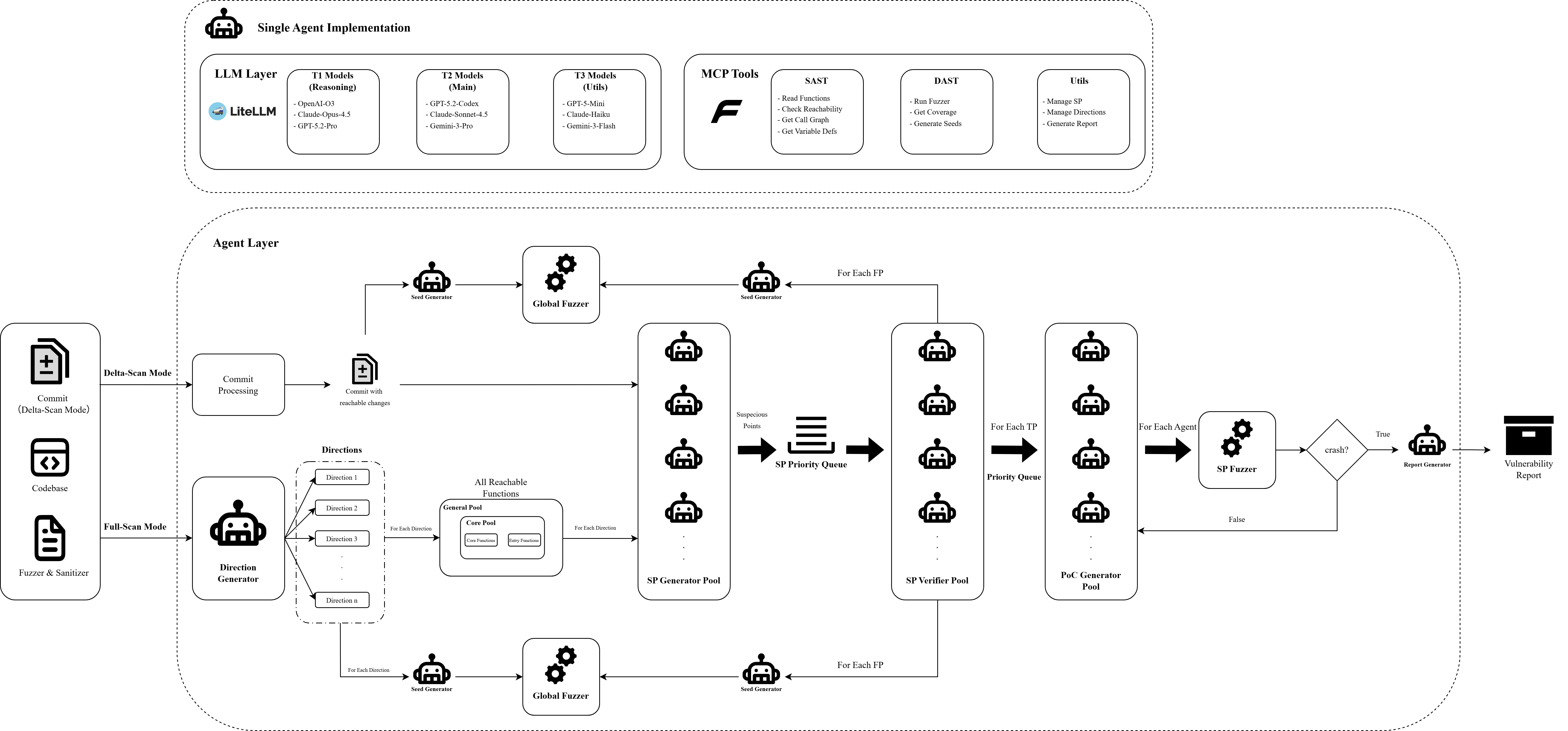}
\caption{Per-worker pipeline. Upper: single agent implementation with LLM tiers (T1 reasoning, T2 main, T3 utils) and MCP tools (SAST, DAST, Utils). Lower: agent layer workflow showing Full-Scan mode (direction-based) and Delta-Scan mode (commit-based) converging into the SP generation, verification, and PoC generation pipeline.}
\label{fig:worker-pipeline}
\end{figure*}

Each worker executes the agent pipeline according to its scan mode (Figure~\ref{fig:worker-pipeline}): Full-Scan mode performs comprehensive analysis via direction-based search, while Delta-Scan mode focuses on changed functions from a commit.

\subsection{Logic-Driven Search (Full-Scan Mode)}
\label{sec:direction}

In Full-Scan mode (lower-left of Figure~\ref{fig:worker-pipeline}), a worker analyzes the entire codebase reachable by its assigned fuzzer-sanitizer pair. The Direction Generator reads the fuzzer and codebase to generate \textit{directions}—each representing a business feature. The SP Generator then analyzes functions according to a tiered scheduling strategy, producing SPs that are subsequently verified. Finally, verified SPs undergo PoC generation with dual-layer fuzzing support.

\subsubsection{Direction}
\label{sec:direction-def}

A \textit{direction} represents a logical feature of the codebase. We organize search by directions rather than vulnerability patterns (e.g., ``buffer overflow'', ``use-after-free'') based on a key insight: fuzzers are designed to test \textit{features}, not to find specific bug types. This semantic grouping extracts related functions, enabling subsequent agents to focus on cohesive code regions. LLMs excel at code semantic understanding, making them well-suited for this task.

\begin{figure}[H]
\begin{tcolorbox}[colback=blue!5, colframe=blue!40, title=Direction Example (libpng~\cite{libpng})]
\small\raggedright
\textbf{name}: PNG chunk parsing\\[2pt]
\textbf{entry\_functions}: [\texttt{read\_chunk\_header}, ...]\\[2pt]
\textbf{core\_functions}: [\texttt{handle\_IHDR}, \texttt{handle\_PLTE}, ...]\\[2pt]
\textbf{risk\_level}: high/medium/low\\[2pt]
\textbf{risk\_reason}: Parses untrusted chunk data with variable-length fields
\end{tcolorbox}
\caption{Example direction for libpng~\cite{libpng}. Each direction includes a business feature name, entry/core functions defining analysis scope, and risk level for prioritization.}
\label{fig:direction-example}
\end{figure}

The Direction Generator (prompt in Figure~\ref{fig:direction-prompt}) first analyzes the fuzzer's source code to understand what it tests, then explores the codebase via MCP tools before invoking \texttt{create\_direction} to generate at most 5 directions per worker (Figure~\ref{fig:direction-example}). For each direction, it produces: a \texttt{name} describing the business feature, \texttt{entry\_functions} where this feature's logic begins, \texttt{core\_functions} implementing the logic, and \texttt{risk\_level} (high/medium/low) for prioritization.

\subsubsection{Priority Scheduling}
\label{sec:priority}

From each direction, we extract functions into two pools based on their role:
\begin{itemize}
\item \textbf{Core Pool}: Entry functions and core functions identified by the Direction Generator—high-priority targets.
\item \textbf{General Pool}: Other reachable functions within the direction—analyzed on a best-effort basis.
\end{itemize}

We adopt a function-centric analysis strategy: each function is analyzed individually. When given multiple functions at once, LLMs tend to selectively focus on some while overlooking others. By analyzing one function at a time—while providing callers, callees, and MCP tools to explore context—we ensure the agent examines all suspicious patterns within each function thoroughly.

Functions are scheduled according to a priority matrix (Table~\ref{tab:priority}) combining pool membership and global analysis status.

\begin{table}[t]
  \centering
  \caption{Priority scheduling matrix combining pool membership and analysis status.}
  \label{tab:priority}
  \small
  \setlength{\tabcolsep}{6pt}
  \begin{threeparttable}
    \begin{tabularx}{\columnwidth}{@{}l>{\centering\arraybackslash}X>{\centering\arraybackslash}X@{}}
      \toprule
      \textbf{Pool} & \textbf{Unanalyzed} & \textbf{Analyzed}\tnote{\dag} \\
      \midrule
      Core    & Priority 1\tnote{*} & Priority 2 \\
      General & Priority 3          & Priority 4 \\
      \bottomrule
    \end{tabularx}
    \begin{tablenotes}[flushleft]
      \footnotesize
      \item[*] Highest priority.
      \item[\dag] Examined by other directions but may yield different insights.
    \end{tablenotes}
  \end{threeparttable}
\end{table}

Priority scheduling improves search efficiency. Since a function may appear in the core pools of multiple directions, this design prevents the same function from being analyzed repeatedly by different directions within a short period, maximizing the breadth of function coverage.

\subsubsection{SP Generation \& Deduplication}
\label{sec:spgen}

The SP Generator (prompt in Figure~\ref{fig:sp-prompt}) performs initial screening on each function. It adopts a high-recall strategy: report potential issues liberally, as a dedicated Verifier will filter false positives later. The generator is sanitizer-aware, focusing on patterns detectable by the current worker's sanitizer (e.g., buffer operations for AddressSanitizer, uninitialized memory for MemorySanitizer).

Since the same vulnerability may be discovered multiple times—by different directions within a worker, by different workers analyzing overlapping code, or by the same generator in different iterations—we employ an SP Deduplicator (T3 model) to identify duplicates. The deduplicator compares each new SP against existing SPs in the database, checking function name, vulnerability type, and description similarity. Duplicates are merged rather than discarded: source information is combined, and the higher confidence score is retained.

\subsubsection{SP Verification}
\label{sec:spverify}

The SP Verifier (prompt in Figure~\ref{fig:spverify-prompt}) performs in-depth analysis on each SP. As the most capable agent in the pipeline, it has full access to code analysis tools and employs a T3 utility model for intelligent context compression when tracing cross-function data flows.

The Verifier examines three aspects:
\begin{itemize}
\item \textbf{Reachability}: Not just function reachability, but whether fuzzer input can reach the specific vulnerable code path.
\item \textbf{Safety boundaries}: Traces context to verify whether existing protections (bounds checks, input validation) are actually correct—they often contain subtle flaws.
\item \textbf{Accuracy}: Validates whether the SP description matches the actual code. If the location is correct but the description is wrong, the Verifier corrects it rather than rejecting the SP.
\end{itemize}

The design principle remains conservative: PoC failure is cheap, but missing a real bug is expensive. The Verifier only rejects SPs when 100\% certain they are false positives.

The Verifier classifies each SP as either TP (true positive candidate) or FP (false positive). TPs enter a priority queue ordered by importance and score, proceeding to PoC generation with basic guidance. FPs are handled by the fuzzing layer (Section~\ref{sec:fuzzing}).

\subsubsection{PoC Generation and Reporting}
\label{sec:pocgen}

The PoC Generator (prompt in Figure~\ref{fig:poc-prompt}) processes SPs from the priority queue, iteratively crafting inputs to trigger each vulnerability. For each attempt, the generator produces Python code that outputs multiple blob variants—using code generation rather than direct byte output ensures format correctness and enables structured exploration of the input space. Each blob is verified individually: on success, the SP is confirmed as a real vulnerability; on failure, the generator receives execution output hints and the blob is added to the SP Fuzzer corpus (Section~\ref{sec:fuzzing}) for mutation.

After several failed attempts, dynamic tracing becomes available, revealing which functions were executed and where the path diverged from the target. This progressive strategy balances token cost: dynamic tracing provides more information but consumes more context, so we reserve it for difficult cases.

The generator iterates until a crash is confirmed or configurable attempt limits are reached. When an SP is reachable by multiple fuzzers, each fuzzer attempts PoC generation independently, increasing success probability through different entry paths.

Upon successful crash, a Report Agent analyzes the crash trace to refine the SP description, then generates a structured JSON output along with a human-readable vulnerability report. The report can be directly submitted to GitHub Issues or Security Advisories.

\subsection{Delta-Scan Processing}
\label{sec:delta}

In Delta-Scan mode, FuzzingBrain V2 analyzes code changes between two versions (e.g., a commit) rather than the entire codebase. The system parses the diff to extract reachable modified functions—these functions replace directions as the analysis scope. We preserve commit messages and comments in the diff context, as they may help agents understand the intent behind code changes; if irrelevant, the Context Compressor removes them automatically. The subsequent pipeline (SP Generation, SP Verification, PoC Generation) follows the same logic as Full-Scan.

\subsection{Dual-layer Fuzzing}
\label{sec:fuzzing}

FuzzingBrain V2 employs two fuzzer layers (Figure~\ref{fig:worker-pipeline}): Global Fuzzer for breadth and SP Fuzzer for depth.

\smallskip
\noindent\textbf{Global Fuzzer.} Runs continuously in the background for broad exploration. The Seed Generator produces seeds from two sources: (1) directions—when a direction is created, the generator analyzes the fuzzer entry point and core functions to produce targeted seeds; (2) false positives—SPs that the Verifier determined to be non-exploitable are still ``near misses'' where the LLM identified suspicious patterns, so we generate seeds to explore these code regions. Crashes discovered by background fuzzer execution (either Global or SP Fuzzer) are directly packaged as vulnerability reports.

\smallskip
\noindent\textbf{SP Fuzzer.} Activated for each SP during PoC generation, operating in two modes: (1) verification mode—executes each blob generated by the PoC Generator to check if it triggers a crash; (2) background mode—failed PoC attempts are added to the corpus for mutation while the LLM reasons about the next attempt. This parallel execution ensures ``the fuzzer mutates while the LLM thinks.''

\section{Implementation}
\label{sec:impl}

We implement FuzzingBrain V2 in Python, refactoring and extending our prior work~\cite{fuzzingbrain} with a distributed worker architecture. Tasks are dispatched through a Redis message queue to parallel worker processes, each handling one fuzzer-sanitizer pair. MongoDB stores task states, SPs, and vulnerability reports. A separate Analyzer Server process provides code analysis capabilities (call graphs, function extraction, reachability queries) to agents via RPC.

\textbf{Agent Implementation.} All agents inherit from a unified BaseAgent class that implements the MCP tool-calling loop: the agent sends a request to the LLM, receives tool calls, executes them via MCP, and returns results to the LLM until it stops. Each agent instance receives an isolated MCP server (via a factory pattern) to prevent state interference during parallel execution.

\textbf{LLM Integration.} As shown in Table~\ref{tab:agents}, we use a three-tier model configuration with fallback chains: if the primary model fails (rate limit, timeout, or service unavailable), the system automatically retries with alternative models in the same tier. Temperature is set to 0.1 for near-deterministic outputs.

\textbf{Fuzzing Integration.} Both Global and SP Fuzzers use libFuzzer with sanitizer instrumentation (ASan, MSan, UBSan). The system monitors fuzzer output for crashes and automatically triggers PoC verification.

\textbf{Evaluation Infrastructure.} Real-time monitoring tracks token consumption, execution time, tool invocations, and agent status by querying database state.

\section{Evaluation}
\label{sec:eval}

We evaluate FuzzingBrain V2 on the AIxCC 2025 Final Competition dataset and real-world open-source projects. Our evaluation answers the following research questions:

\begin{itemize}
\item \textbf{RQ1 (Effectiveness and Efficiency):} How effective and efficient is FuzzingBrain V2 at discovering vulnerabilities compared to state-of-the-art approaches?
\item \textbf{RQ2 (Complex Case Analysis):} How does FuzzingBrain V2 perform on vulnerabilities with deep call chains and complex dependencies?
\item \textbf{RQ3 (Ablation Study):} How do the different components of FuzzingBrain V2 contribute to its overall effectiveness?
\item \textbf{RQ4 (Zero-day Discovery):} Can FuzzingBrain V2 discover previously unknown vulnerabilities in real-world open-source projects?
\end{itemize}

\subsection{Experimental Setup}

\textbf{Dataset.} We evaluate on the C/C++ portion of the AIxCC 2025 Final Challenge dataset (AFC), which contains 40 vulnerabilities across 12 open-source projects including curl, Wireshark, systemd, and libxml2. The dataset includes 20 Full-scan challenges and 20 Delta-scan challenges. Each vulnerability is verified through sanitizer-detected crashes, ensuring ground truth correctness.

\textbf{Baselines.} We compare against seven baselines: (1) six AIxCC finalist teams: Team Atlanta (AT, 1st place), Trail of Bits (TB, 2nd), Theori (TI, 3rd), FuzzingBrain V1 (FB, 4th), Shellphish (SP, 5th), and 42-beyond-bugs (42, 6th); (2) Claude Code (CC, Claude-Opus-4.5), a state-of-the-art general-purpose coding agent. For CC, we provide the ground-truth fuzzer-sanitizer pair and allow up to 2 hours of human-in-the-loop interaction per challenge to guide the analysis.

\textbf{Configuration.} Each AIxCC challenge specifies multiple fuzzers and sanitizers, and competition teams run parallel workers across all combinations. For our evaluation, we run FuzzingBrain V2 with only the ground-truth fuzzer-sanitizer pair for each challenge, simulating a single-worker configuration. This setup isolates the system's detection capability from parallelization overhead. We set limits of 120 minutes and \$150 for Delta-scan, and 240 minutes and \$400 for Full-scan challenges. We fix model tiers as T1: Claude Opus 4.5, T2: Claude Sonnet 4.5, T3: Claude Haiku 4.5. All experiments run on a server with Intel Xeon Platinum 8272CL CPU (48 cores) and 94GB RAM.

\subsection{RQ1: Effectiveness and Efficiency}

\begin{figure*}[t]
\centering
\includegraphics[width=\textwidth]{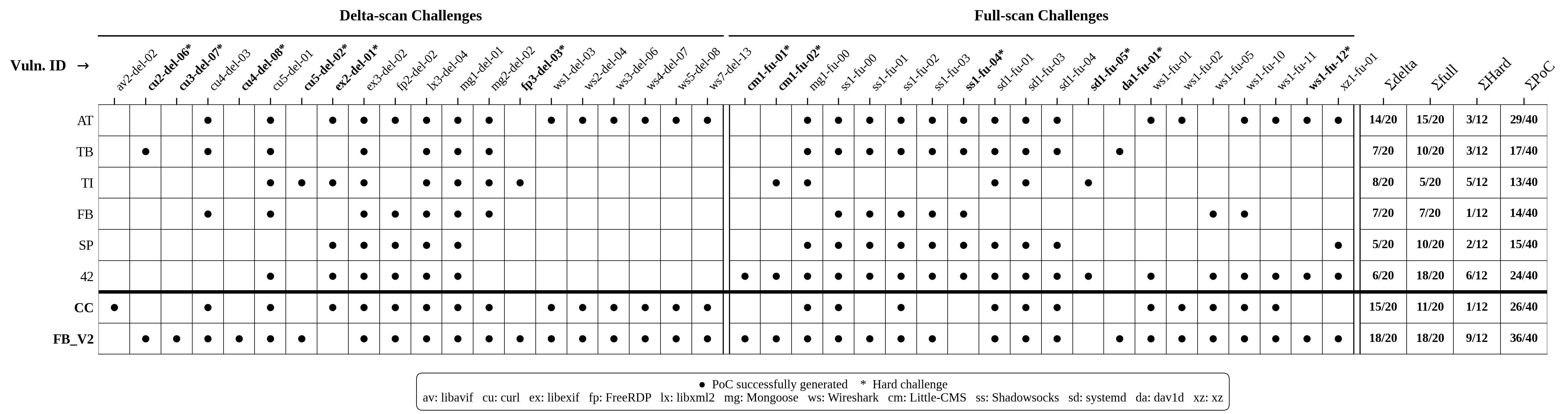}
\caption{Vulnerability discovery results on the AFC (AIxCC Final Challenge) dataset. Each column represents a challenge (xx-del-xx for Delta-scan, xx-fu-xx for Full-scan). A dot indicates successful PoC generation. Challenges marked with * are high-difficulty vulnerabilities that most teams failed to find. The thick horizontal line separates competition teams (above) from our experimental baselines (below). $\Sigma$delta/$\Sigma$full show per-mode totals, $\Sigma$Hard counts high-difficulty challenges solved, and $\Sigma$PoC shows overall totals.}
\label{fig:effectiveness}
\end{figure*}

Figure~\ref{fig:effectiveness} shows the vulnerability discovery results across all 40 AIxCC challenges. FuzzingBrain V2 discovers 36 out of 40 vulnerabilities, ranking first among all systems. This is 7 more than the AIxCC champion Team Atlanta (29) and 22 more than FuzzingBrain V1 (14), representing a 157\% improvement over our previous system.

Notably, FuzzingBrain V2 solves 9 out of 12 hard challenges---vulnerabilities that most competition teams failed to discover (see Section~\ref{sec:rq2} for hard challenge definition). Claude Code (CC), despite being a general-purpose coding agent, achieves 26 vulnerabilities, demonstrating LLM potential but also the benefit of domain-specific optimization.

Table~\ref{tab:detailed-results} shows detailed per-challenge metrics. For each challenge, SP$_{tot}$ and SP$_{ded}$ denote the total and deduplicated suspicious points generated; TP$_v$ and FP indicate how the Verifier classified them; TP$_a$ shows the cumulative discovery count; and By indicates whether the vulnerability was found by the Global Fuzzer (G) or SP-targeted Fuzzer (S).

The SP pipeline generated 1,277 suspicious points (1,030 after deduplication). The Verifier classified 96 as TP$_v$ and 938 as FP. Of the 36 discovered vulnerabilities, 35 were correctly classified as TP$_v$ (recall 97.2\%); one (ws1-fu-12) was misclassified as FP but still discovered through FP-generated seeds. This validates our design choice to prioritize recall---false positives are eliminated by the PoV Generator when no crash is triggered, but missed vulnerabilities are hard to recover.

\textbf{Efficiency.} Successful Delta-scan challenges average 12 minutes and \$19.4, while Full-scan averages 18 minutes and \$35.2. The four failed challenges (av2-del-02, ex2-del-01, sd1-fu-05, ss1-fu-04) consumed significant resources before timeout, indicating complex vulnerabilities requiring deeper analysis. Total cost across all 40 challenges was \$1,785.60 with 526M tokens.

\begin{table*}[!ht]
\centering
\caption{Detailed FuzzingBrain V2 results on AFC dataset. SP = Suspicious Points (Total/Deduplicated). TP$_v$ = verifier-confirmed TPs. TP$_a$ = actual TPs. By = discovery method (G=Global Fuzzer, S=SP Fuzzer).}
\label{tab:detailed-results}
\setlength{\tabcolsep}{6pt}
\begin{tabular}{lllcccccccccc}
\toprule
\textbf{Challenge} & \textbf{San.} & \textbf{Type} & \textbf{PoC} & \textbf{By} & \textbf{SP$_{tot}$} & \textbf{SP$_{ded}$} & \textbf{TP$_v$} & \textbf{TP$_a$} & \textbf{FP} & \textbf{Time} & \textbf{Tok.(K)} & \textbf{Cost} \\
\midrule
\multicolumn{13}{l}{\textit{Delta-scan Challenges}} \\
\midrule
av2-del-02 & ASAN & OOB-W & \xmark & -- & 3 & 3 & 3 & 0 & 0 & T/O & 44,363 & \$135.05 \\
cu2-del-06 & ASAN & NPD & \cmark & S & 1 & 1 & 1 & 1 & 0 & 3m & 4,663 & \$23.76 \\
cu3-del-07 & ASAN & NPD & \cmark & S & 10 & 10 & 5 & 1 & 5 & 1h7m & 31,437 & \$110.77 \\
cu4-del-03 & ASAN & Stack-OF & \cmark & G & 5 & 5 & 2 & 1 & 3 & 2m & 4,358 & \$13.34 \\
cu4-del-08 & ASAN & NPD & \cmark & S & 5 & 5 & 4 & 1 & 1 & 1h40m & 42,726 & \$130.32 \\
cu5-del-01 & ASAN & Fmt-Str & \cmark & S & 6 & 6 & 5 & 1 & 1 & 12m & 9,926 & \$41.23 \\
cu5-del-02 & ASAN & OOB-W & \cmark & S & 1 & 1 & 1 & 1 & 0 & 31m & 9,142 & \$27.79 \\
ex2-del-01 & ASAN & Heap-OF & \xmark & -- & 4 & 3 & 2 & 1 & 2 & T/O & 38,421 & \$118.62 \\
ex3-del-02 & ASAN & Stack-OF & \cmark & G & 0 & 0 & 1 & 1 & 0 & 51s & 57 & \$1.70 \\
fp2-del-02 & ASAN & Heap-OF & \cmark & S & 3 & 3 & 3 & 1 & 0 & 10m & 1,123 & \$3.71 \\
fp3-del-03 & ASAN & Arb-W & \cmark & S & 3 & 3 & 3 & 1 & 0 & 15m & 4,884 & \$14.66 \\
lx3-del-04 & ASAN & Heap-OF & \cmark & G & 4 & 4 & 2 & 1 & 2 & 3m & 4,645 & \$22.01 \\
mg1-del-01 & ASAN & Heap-OF & \cmark & G & 0 & 0 & 1 & 1 & 0 & 60s & 92 & \$2.80 \\
mg2-del-02 & ASAN & Heap-OF & \cmark & S & 1 & 1 & 1 & 1 & 0 & 11m & 1,759 & \$5.37 \\
ws1-del-03 & ASAN & Stack-OF & \cmark & G & 1 & 1 & 1 & 1 & 0 & 1s & 55 & \$0.50 \\
ws2-del-04 & ASAN & Stack-OF & \cmark & G & 1 & 1 & 1 & 1 & 0 & 14s & 518 & \$2.18 \\
ws3-del-06 & ASAN & Heap-OF & \cmark & G & 1 & 1 & 1 & 1 & 0 & 17s & 454 & \$1.55 \\
ws4-del-07 & ASAN & Global-OF & \cmark & G & 1 & 1 & 1 & 1 & 0 & 23s & 1,941 & \$6.70 \\
ws5-del-08 & ASAN & Global-OF & \cmark & G & 2 & 1 & 1 & 1 & 1 & 18s & 2,156 & \$7.24 \\
ws7-del-13 & ASAN & Heap-OF & \cmark & G & 1 & 1 & 1 & 1 & 0 & 8s & 487 & \$1.82 \\
\midrule
\multicolumn{13}{l}{\textit{Full-scan Challenges}} \\
\midrule
cm1-fu-01* & ASAN & NPD & \cmark & S & 133 & 48 & 7 & 1 & 36 & 1h50m & 43,984 & \$150.38 \\
cm1-fu-02 & ASAN & Stack-OF & \cmark & S & 171 & 62 & 12 & 2 & 45 & 2h10m & 65,946 & \$227.20 \\
da1-fu-01* & UBSAN & Int-OF & \cmark & S & 54 & 12 & 3 & 1 & 5 & 48m & 7,960 & \$29.07 \\
mg1-fu-00* & ASAN & Heap-OF & \cmark & G & 10 & 10 & 4 & 1 & 6 & 51m & 18,119 & \$56.78 \\
sd1-fu-01* & ASAN & Heap-OF & \cmark & S & 9 & 6 & 3 & 2 & 2 & 6m & 3,739 & \$14.89 \\
sd1-fu-03 & ASAN & NPD & \cmark & G & 14 & 10 & 5 & 3 & 4 & 11m & 4,619 & \$17.89 \\
sd1-fu-04 & ASAN & Stack-OF & \cmark & G & 5 & 3 & 2 & 1 & 1 & 3m & 4,713 & \$18.22 \\
sd1-fu-05 & ASAN & DblFree & \xmark & -- & 487 & 465 & 12 & 4 & 453 & T/O & 113,373 & \$379.38 \\
ss1-fu-00* & ASAN & Heap-OF & \cmark & G & 8 & 5 & 2 & 1 & 2 & 2m & 1,850 & \$6.12 \\
ss1-fu-01 & ASAN & Heap-OF & \cmark & S & 22 & 15 & 5 & 3 & 8 & 7m & 5,680 & \$18.75 \\
ss1-fu-02 & ASAN & Heap-OF & \cmark & S & 22 & 15 & 5 & 4 & 8 & 7m & 5,720 & \$18.92 \\
ss1-fu-03 & ASAN & Heap-OF & \cmark & G & 14 & 9 & 3 & 2 & 5 & 4m & 3,420 & \$11.28 \\
ss1-fu-04 & ASAN & Heap-OF & \xmark & -- & 438 & 412 & 14 & 5 & 398 & T/O & 98,752 & \$348.65 \\
ws1-fu-01* & ASAN & Stack-OF & \cmark & G & 5 & 5 & 1 & 2 & 0 & 3m & 2,148 & \$10.24 \\
ws1-fu-02 & ASAN & UAF & \cmark & G & 5 & 5 & 1 & 5 & 0 & 5m & 2,716 & \$13.52 \\
ws1-fu-05 & ASAN & Fmt-Str & \cmark & G & 5 & 5 & 1 & 3 & 0 & 4m & 2,168 & \$10.78 \\
ws1-fu-10 & ASAN & Heap-OF & \cmark & G & 5 & 5 & 1 & 1 & 0 & 3m & 2,070 & \$10.08 \\
ws1-fu-11 & ASAN & Buf-OF & \cmark & G & 5 & 5 & 1 & 4 & 0 & 4m & 2,271 & \$11.35 \\
ws1-fu-12 & ASAN & NPD & \cmark & S & 8 & 6 & 5 & 6 & 1 & 10m & 5,842 & \$28.65 \\
xz1-fu-01* & ASAN & UAF & \cmark & G & 56 & 20 & 6 & 1 & 14 & 40m & 13,204 & \$44.75 \\
\midrule
\multicolumn{3}{l}{\textbf{Total}} & \textbf{36/40} & & & & & & & & \textbf{526,403} & \textbf{\$1,785.60} \\
\bottomrule
\multicolumn{13}{l}{\footnotesize *First vulnerability of each Full-scan challenge; same prefix (e.g., cm1-fu-01, cm1-fu-02) = same challenge, shared SP analysis. Metrics are cumulative} \\
\multicolumn{13}{l}{\footnotesize \hspace{0.5em} snapshots at discovery.} \\
\end{tabular}
\end{table*}

\subsection{RQ2: Complex Case Analysis}
\label{sec:rq2}

As shown in Figure~\ref{fig:effectiveness}, we manually reviewed all 40 challenges and identified 12 \textit{hard challenges} based on the following criteria: (1) deep call chains with complex dependencies; (2) vulnerabilities masked by other bugs, where triggering the target vulnerability is blocked by earlier crashes; (3) requiring sequential PoC execution to reach the vulnerable code path. We excluded most Wireshark (ws), systemd (sd), and libavif (av) vulnerabilities, as their detection failures stemmed from build process difficulties rather than analysis complexity.

FuzzingBrain V2 discovers 9 out of 12 hard challenges (75\%), significantly outperforming Team Atlanta (5/12, 42\%). Notably, all 9 discovered hard challenges were found by the SP Fuzzer, not the Global Fuzzer. This highlights the importance of deep exploration capability: while the Global Fuzzer relies on coverage-guided mutation, the SP Fuzzer combines static analysis with dynamic feedback, enabling it to reason about complex call chains and craft inputs that reach deeply nested vulnerable code.

\begin{figure}[!ht]
\centering
\includegraphics[width=\columnwidth]{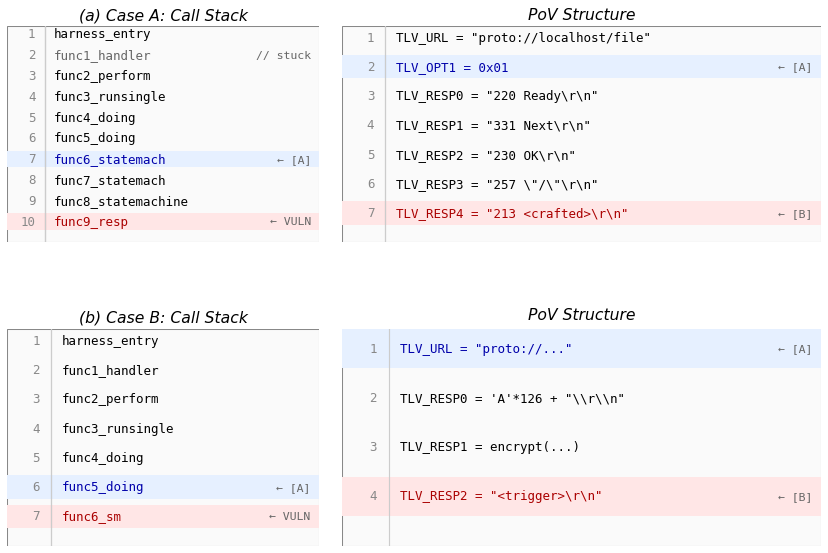}
\vspace{-2mm}
\caption{PoV requirements for two hard challenges. (a) Leap second OOB requires historical timestamp. (b) Type confusion requires protocol bit collision and AES encryption. Both require precise multi-field coordination that random fuzzing cannot achieve.}
\label{fig:pov-structure}
\end{figure}

\begin{figure}[!ht]
\centering
\includegraphics[width=\columnwidth]{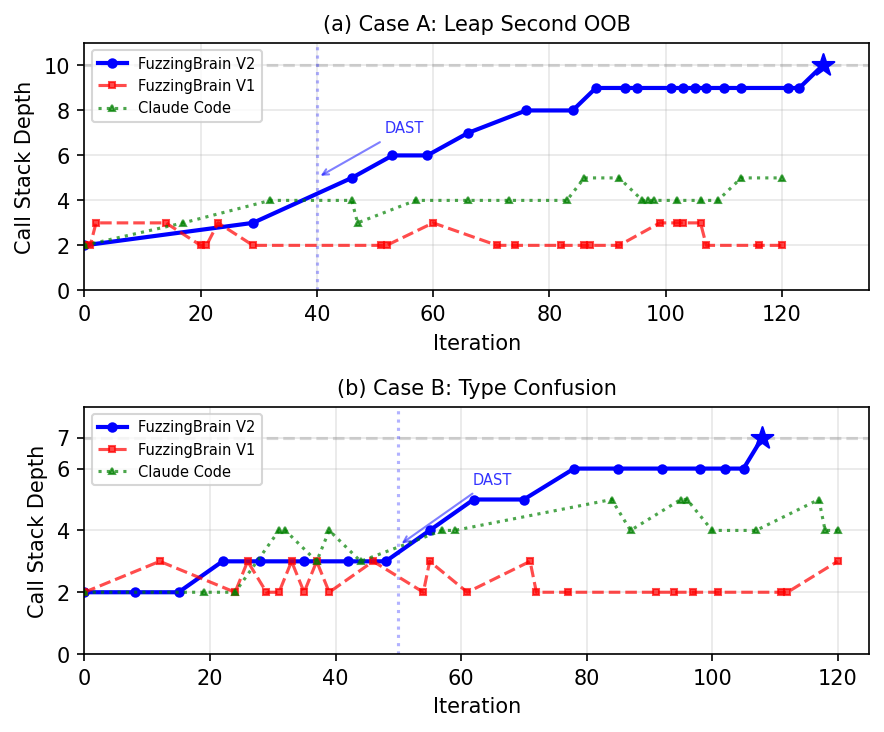}
\vspace{-2mm}
\caption{PoV generation progress. FuzzingBrain V2 progressively deepens while FBv1 and CC remain stuck at shallow depths.}
\label{fig:case-study}
\vspace{2mm}
\end{figure}

We present two case studies that illustrate FuzzingBrain V2's capability to discover vulnerabilities requiring deep domain knowledge. We select Case A (discovered by 2 of 9 teams) and Case B (discovered only by FuzzingBrain V2). We omit a third case from this analysis---though also uniquely discovered by FuzzingBrain V2---because its vulnerability type and PoV structure are similar to Case B. Figure~\ref{fig:pov-structure} shows the call stack and PoV structure for each case; Figure~\ref{fig:case-study} shows the exploration progress over iterations.

\textbf{Case A (Leap Second OOB).} This vulnerability in a protocol's timestamp handler triggers only when parsing a response containing a leap second (seconds=60). The bug lies at call depth 10 in a response handler function, where a timestamp lookup array is accessed with index 60 in a 60-element array. The challenge: most leap second timestamps are rejected by the date validation function; only specific historical leap seconds bypass validation. FuzzingBrain V2's agent discovered this through iterative hypothesis testing---after DAST tools revealed that the vulnerable function was not being reached, the agent systematically tried different historical leap seconds until finding one that worked.

\textbf{Case B (Type Confusion).} This vulnerability exploits a protocol bit collision: two protocol constants share the same bit value, allowing an unexpected protocol handler to be reached through the fuzzer. The bug at call depth 7 dereferences an enum value as a pointer. The challenge: reaching the vulnerable code requires (1) discovering the bit collision, (2) implementing correct AES-256-CBC encryption with hardcoded key/IV found in source code, and (3) sending an exact trigger string. The agent reversed the encryption scheme by reading the source code and correctly implemented the three-phase protocol handshake.

Both cases demonstrate that FuzzingBrain V2 can reason about complex constraints that defeat random fuzzing. Figure~\ref{fig:case-study} shows the exploration progress over iterations. In Case A, FuzzingBrain V2 requires 127 iterations to reach depth 10; DAST enables a breakthrough after iteration 40 by revealing that the vulnerable function was not being reached. In Case B, FuzzingBrain V2 requires 108 iterations to reach depth 7; DAST at iteration 50 helps the agent discover the protocol bit collision and reverse the AES encryption. In both cases, FBv1 and Claude Code remain stuck at shallow depths (2--5), unable to reason about the complex constraints required to reach the vulnerable code.

These case studies demonstrate FuzzingBrain V2's deep vulnerability detection capability. Through MCP-based tool invocation, agents autonomously leverage both static analysis (to understand code structure, trace data flow, and identify constraints) and dynamic analysis (to obtain concrete execution feedback and verify hypotheses). This combination enables systematic exploration of complex input constraints that neither traditional fuzzing nor standalone LLM agents can achieve.

The three unsolved hard challenges reveal FuzzingBrain V2's limitations: one requires simultaneous multi-input attacks unsupported by our architecture; one triggered a similar shallower bug causing premature worker termination; and one involves an implicit state machine our static analysis could not reconstruct.

\subsection{RQ3: Ablation Study}

\begin{figure*}[!ht]
\centering
\includegraphics[width=\textwidth]{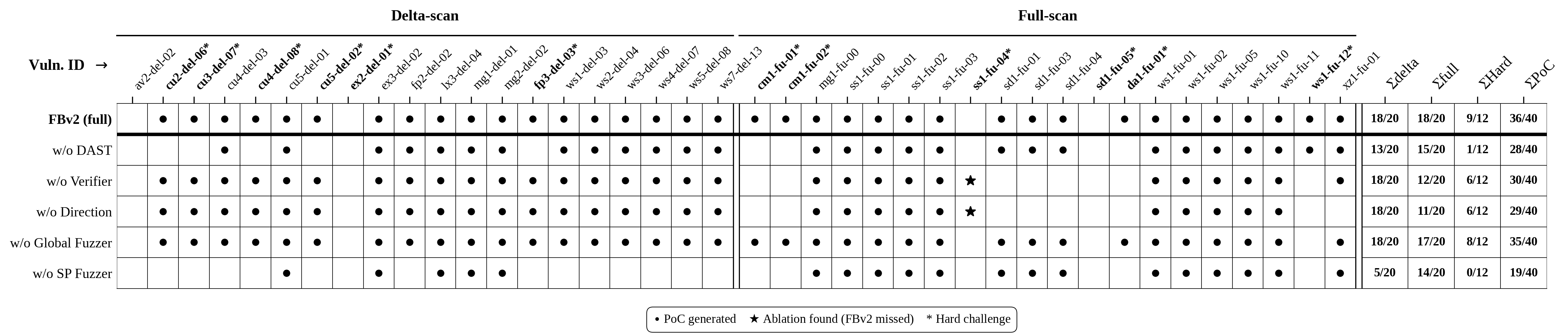}
\caption{Ablation study results. Stars indicate vulnerabilities found by ablation configurations but missed by FuzzingBrain V2 (full).}
\label{fig:ablation}
\end{figure*}

To understand the contribution of each component, we conduct ablation experiments with the same resource limits as RQ1. Figure~\ref{fig:ablation} shows the results; Table~\ref{tab:ablation-time} shows the average time per successful discovery.

Removing dynamic analysis tools (w/o DAST) reduces detection from 36 to 28, with hard challenge coverage dropping from 9/12 to 1/12. Average time increases (Delta: 25 vs 12 min, Full: 27 vs 18 min) while effectiveness drops significantly. Without execution feedback, agents cannot verify whether target code paths are reached, leaving them stuck at shallow depths.

Without verification (w/o Verifier), detection drops from 36 to 30. More critically, average time increases 3$\times$ for Delta-scan (38 vs 12 min), and most Full-scan tasks timeout due to SP accumulation. Interestingly, one vulnerability (ss1-fu-04) was found without the Verifier but missed by the full system---likely due to bypassing a blocking vulnerability that previously prevented reaching the target.

Removing directions (w/o Direction) reduces detection from 36 to 29. Delta-scan time remains unchanged (12 min) since Delta-scan does not use the direction generation module. However, Full-scan time increases nearly 3$\times$ (48 vs 18 min) due to no priority scheduling---without Core Pool and General Pool, low-priority functions consume PoC Generator resources and cause timeouts.

Without the Global Fuzzer (w/o Global Fuzzer), FuzzingBrain V2 achieves 35/40 with 8/12 hard challenges but takes longer (Delta: 24 vs 12 min, Full: 31 vs 18 min). Without the SP Fuzzer (w/o SP Fuzzer), the system is fast (2 min) but achieves only 19/40 with zero hard challenges. This confirms that LLM-guided SP exploration is essential for complex vulnerabilities, while background fuzzing provides broad but shallow coverage.

\begin{table}[H]
\centering
\caption{Average time per successful vulnerability discovery.}
\label{tab:ablation-time}
\small
\begin{threeparttable}
\begin{tabular}{lcc}
\toprule
\textbf{Configuration} & \textbf{Delta (min)} & \textbf{Full (min)} \\
\midrule
FuzzingBrain V2 (full) & 12 & 18 \\
w/o DAST & 25 & 27 \\
w/o Verifier & 38 & T/O\tnote{*} \\
w/o Direction & 12 & 48 \\
w/o Global Fuzzer & 24 & 31 \\
w/o SP Fuzzer & 2 & 2 \\
\bottomrule
\end{tabular}
\begin{tablenotes}
\footnotesize
\item[*] Most Full-scan tasks timeout due to SP accumulation; not statistically meaningful.
\end{tablenotes}
\end{threeparttable}
\end{table}

\subsection{RQ4: Zero-day Discovery}

To evaluate FuzzingBrain V2's real-world applicability, we deployed the system on OSS-Fuzz projects and collaborative targets from our lab partnerships. Each project was allocated a \$200 budget, adjusted based on the number of fuzzers available. All targets are either directly from OSS-Fuzz or adapted to the OSS-Fuzz infrastructure.

\begin{table}[H]
\centering
\caption{Zero-day vulnerabilities discovered by FuzzingBrain V2.}
\label{tab:zero-day}
\small
\setlength{\tabcolsep}{3pt}
\begin{threeparttable}
\begin{tabular}{llcccc}
\toprule
\textbf{Project} & \textbf{Lang.} & \textbf{Found} & \textbf{Submitted} & \textbf{Confirmed}\tnote{*} & \textbf{Fixed} \\
\midrule
OpenPrint CUPS & C & 6 & 6 & 6 & 6 \\
bluez & C & 5 & 5 & 0 & 0 \\
fwupd & C & 4 & 4 & 4 & 4 \\
upx & C++ & 4 & 4 & 4 & 4 \\
avro & Java/C & 3 & 3 & 3 & 3 \\
pdfbox & Java & 3 & 3 & 0 & 0 \\
imagemagick & C & 2 & 2 & 2 & 2 \\
V2xHub & C++ & 2 & 2 & 0 & 0 \\
busybox & C & 2 & 2 & 0 & 0 \\
cjson & C & 1 & 1 & 0 & 0 \\
flatbuffers & C++ & 1 & 1 & 0 & 0 \\
jq & C & 1 & 1 & 1 & 0 \\
MongoDB & C++ & 1 & 1 & 1 & 0 \\
mongoose & C & 1 & 1 & 1 & 1 \\
njs & C & 1 & 1 & 1 & 1 \\
openldap & C & 1 & 1 & 0 & 0 \\
paddle & C++ & 1 & 1 & 1 & 1 \\
simdutf & C++ & 1 & 1 & 1 & 1 \\
tcpreplay & C & 1 & 1 & 1 & 0 \\
\midrule
\textbf{Total} & -- & \textbf{41} & \textbf{41} & \textbf{26} & \textbf{23} \\
\bottomrule
\end{tabular}
\begin{tablenotes}
\footnotesize
\item[*] Acknowledged by maintainers with issue tracking.
\end{tablenotes}
\end{threeparttable}
\end{table}

Table~\ref{tab:zero-day} summarizes the results. FuzzingBrain V2 discovered 41 previously unknown vulnerabilities across 19 projects. Our deployment began in early 2026 and is still in its initial stage; as of submission, 26 have been confirmed (acknowledged by maintainers with issue tracking) and 23 have been fixed. Although this paper focuses on C/C++ evaluation, FuzzingBrain V2 can also handle Java projects through OSS-Fuzz's Jazzer integration. While we did not implement Java-specific optimizations (as we did for C), we included Java targets (avro, pdfbox) in our deployment and successfully discovered vulnerabilities.

We did not request CVE assignments from developers, as our collaboration with open-source projects is directly reported to DARPA for verification. Nevertheless, some maintainers proactively applied for CVEs---for example, ImageMagick vulnerabilities were assigned CVE-2026-23874 and CVE-2026-23952.

Notable discoveries include 6 vulnerabilities in OpenPrint CUPS, and 4 each in fwupd and upx. These projects are mature and well-fuzzed, yet FuzzingBrain V2's semantic analysis identified vulnerabilities in code paths that traditional fuzzers had not explored.

\begin{figure}[H]
\centering
\includegraphics[width=0.8\columnwidth]{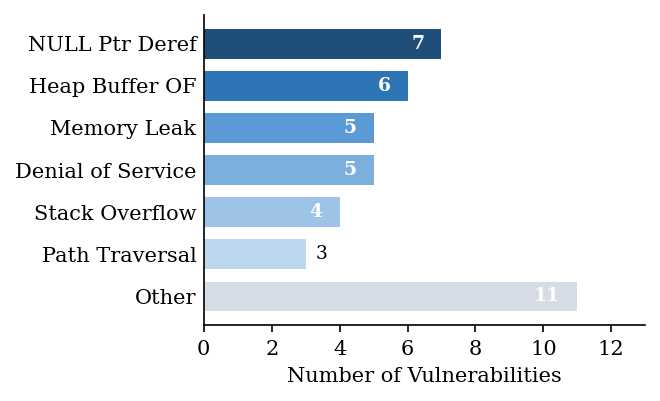}
\caption{Distribution of vulnerability types.}
\label{fig:vuln-types}
\end{figure}

Figure~\ref{fig:vuln-types} shows the distribution of vulnerability types. NULL pointer dereferences (7) and heap buffer overflows (6) are the most common, followed by memory leaks (5) and denial-of-service vulnerabilities (5). This distribution closely mirrors the AFC benchmark dataset, suggesting that FuzzingBrain V2's detection capabilities generalize well to real-world scenarios.

\paragraph{Case Study: Write-Before-Check in Project A.}
Project~A is a mature Unicode processing library (1,700+ stars, no vulnerabilities reported in six months). FuzzingBrain V2 discovered a heap-buffer-overflow caused by a ``write-before-check'' pattern in the UTF-16 to UTF-8 conversion.

\emph{Why traditional fuzzing missed it.} The existing OSS-Fuzz harness targets the standard conversion API, which internally allocates a sufficiently large output buffer. The vulnerable code path is only reachable through the \texttt{\_safe()} variant, which accepts a user-provided buffer with explicit size constraints.

\emph{How FuzzingBrain V2 found it.} The Direction agent flagged \texttt{\_safe()} functions as high-priority, reasoning that safety-critical wrappers often hide edge cases. The SP Fuzzer generated a targeted harness that supplies minimal output buffers. Within 12 minutes, the verifier confirmed a heap overflow: the ASCII fast-path optimization writes output bytes \emph{before} checking buffer bounds, causing overflow when triggered near the boundary.

The maintainers fixed the bug within one week and sent a letter of appreciation.

\section{Discussion}
\label{sec:discussion}

\subsection{Limitations}

Despite achieving 90\% detection rate (36/40) on the AFC benchmark, FuzzingBrain V2 has notable limitations.

\textbf{Multi-input vulnerabilities.} FuzzingBrain V2's architecture assumes a single-input fuzzing model, where one PoC triggers one vulnerability. However, some vulnerabilities require coordinated multi-input attacks---for example, sending a malformed request that corrupts state, followed by a normal request that triggers the crash. Our current SP model cannot represent such temporal dependencies.

\textbf{Implicit state machines.} When program behavior depends on implicit state transitions not reflected in the source code (e.g., protocol handlers with undocumented modes), our static analysis fails to reconstruct the necessary preconditions. The agent may repeatedly generate inputs that are rejected by state validation, unable to discover the correct sequence.

\textbf{Shallow bug masking.} When a shallower bug exists on the path to a deeper vulnerability, FuzzingBrain V2 may terminate exploration prematurely. Although the Verifier attempts to filter duplicate crashes, semantically similar bugs with different root causes can still cause early termination.

\textbf{Build and environment complexity.} Several AFC challenges failed not due to analysis limitations but due to build process difficulties. Complex build systems, missing dependencies, or environment-specific configurations can prevent FuzzingBrain V2 from successfully instrumenting and running the target.

\textbf{Context management.} Our current context compression strategy is simplistic, causing significant information loss during summarization. As a result, agents may repeatedly analyze the same functions across iterations, leading to redundant API calls and increased costs. More sophisticated context management---such as hierarchical summarization or retrieval-augmented memory---could reduce this overhead.

\subsection{Future Work}

\textbf{Multi-language support.} While FuzzingBrain V2 currently focuses on C/C++ with limited Java support through Jazzer, extending to other languages (Rust, Go, Python) would broaden applicability. Each language requires adapting the static analysis pipeline and integrating appropriate sanitizers.

\textbf{Context management.} Adopting retrieval-augmented memory or hierarchical context summarization could significantly reduce redundant analysis and lower operational costs. Techniques from long-context LLM research may also help agents maintain coherent analysis across extended sessions.

\textbf{Patch generation.} A natural extension is automatic patch generation for discovered vulnerabilities. The agent already understands the root cause through PoC generation; generating and validating fixes would complete the vulnerability lifecycle.

\textbf{Binary reproduction.} Currently FuzzingBrain V2 requires source code for instrumentation. Supporting binary-only targets through emulation-based fuzzing (e.g., QEMU, Unicorn) would enable analysis of closed-source software and firmware.

\section{Conclusion}
\label{sec:conclusion}

We presented FuzzingBrain V2, a multi-agent system that combines LLM-driven semantic analysis with coverage-guided fuzzing for automated vulnerability discovery. By introducing the Suspicious Point abstraction, FuzzingBrain V2 enables systematic exploration of potentially vulnerable code while avoiding the infinite search space problem that plagues pure LLM approaches.

On the AFC benchmark, FuzzingBrain V2 detected 36 of 40 vulnerabilities (90\%), including 9 of 12 hard challenges requiring deep semantic understanding. In real-world deployment across 19 OSS-Fuzz projects, FuzzingBrain V2 discovered 41 previously unknown vulnerabilities, of which 26 have been confirmed and 23 fixed by maintainers.

Our results demonstrate that LLM agents can effectively guide vulnerability discovery when grounded by concrete execution feedback and structured by well-defined analysis abstractions. We hope FuzzingBrain V2 inspires further research at the intersection of program analysis and large language models.

\clearpage
\bibliographystyle{plain}
\bibliography{refs}

\appendix
\section{Agent Prompts}
\label{sec:appendix-prompt}

\begin{figure*}[t]
\begin{tcolorbox}[colback=orange!5, colframe=orange!40, title=Direction Generator Prompt (Abridged)]
\small\ttfamily
You are a security architect analyzing a codebase to find vulnerabilities.

\textbf{Background} --- Divide codebase into logical "directions" based on BUSINESS LOGIC for independent analysis.

\textbf{CRITICAL: Understanding Your Constraints} --- FUZZER determines reachability; SANITIZER determines detectability. Static call graph misses function pointer calls.

\textbf{Your Mission}

1. \textbf{Read the fuzzer source code FIRST}
\begin{itemize}
\item Understand what the fuzzer is testing (its PURPOSE)
\item Identify what data format/protocol it processes (its TARGET)
\item List the business functions it exercises (its SCOPE)
\end{itemize}

2. \textbf{Divide by BUSINESS LOGIC, not vulnerability type}
\begin{itemize}
\item Each direction should represent a logical feature or sub-feature
\item Think: "What different things does this code DO?"
\item NOT: "What types of bugs might exist?"
\end{itemize}

3. \textbf{Create directions for each business area}
\begin{itemize}
\item Assign risk levels based on input proximity and complexity
\item Ensure full coverage of reachable functions
\end{itemize}

\textbf{What is a Direction?}

A direction is a logical grouping of functions that handle ONE BUSINESS FEATURE.

\textbf{GOOD direction names} (business logic oriented):
\begin{itemize}
\item Named after WHAT the code DOES (a specific feature or sub-feature)
\item Represents a complete logical unit of functionality
\item Can be understood without security knowledge
\end{itemize}

\textbf{BAD direction names} (DO NOT DO THIS):
\begin{itemize}
\item "Memory Management" (too generic, crosses all features)
\item "Input Parsing" (too vague, every feature parses input)
\item "Buffer Operations" (this is a vulnerability pattern, not a business)
\end{itemize}

\textbf{Security Risk Assessment} --- HIGH: direct untrusted input parsing. MEDIUM: validated data processing. LOW: minimal input dependency.

\textbf{Available Tools} --- get\_function\_source, get\_callers, get\_callees, get\_call\_graph, get\_reachable\_functions, get\_unreached\_functions, search\_code, create\_direction.

\textbf{CRITICAL: Function Pointer Reachability} --- Static analysis cannot track indirect calls. Must actively discover patterns: struct members holding function pointers, callbacks, handler dispatch. These are HIGH VALUE targets.

\textbf{Workflow} --- (1) Read fuzzer source. (2) Get reachable functions. (3) Discover indirect call patterns. (4) Identify business features. (5) Create directions with name, risk\_level, core\_functions, entry\_functions.

\textbf{Important Guidelines} --- Max 5 directions. Full coverage including pointer-reachable functions. Prioritize HIGH RISK first.
\end{tcolorbox}
\caption{Direction Generator Prompt (abridged; full prompt available in supplementary material)}
\label{fig:direction-prompt}
\end{figure*}

\begin{figure*}[t]
\begin{tcolorbox}[colback=orange!5, colframe=orange!40, title=SP Generator Prompt (Abridged)]
\small\ttfamily
You are a vulnerability hunter. Your job is to FIND suspicious code patterns.

\textbf{Your Role: Initial Screening} --- You are the FIRST PASS - an expert Verify Agent will review every SP you create. You don't need to be 100\% certain or fully verify reachability.

\textbf{Key Principle}: It's better to report a potential issue and be wrong, than to miss a real bug because you talked yourself out of it.

\textbf{Your Constraints}
\begin{itemize}
\item \textbf{Fuzzer}: Only code reachable from this fuzzer matters
\item \textbf{Sanitizer}: Only bugs this sanitizer can detect matter
  \begin{itemize}
  \item AddressSanitizer: buffer overflow, OOB, use-after-free, double-free
  \item MemorySanitizer: uninitialized memory read
  \item UndefinedBehaviorSanitizer: integer overflow, null deref, div-by-zero
  \end{itemize}
\end{itemize}

\textbf{When to Create an SP}
\begin{itemize}
\item CREATE: Dangerous pattern + input influence + uncertain protection
\item DON'T skip just because there's a bounds check nearby (it might be wrong)
\end{itemize}

\textbf{Confidence Scores} --- 0.6-1.0: Clear pattern. 0.4-0.6: Suspicious. 0.3-0.4: Worth checking. Only skip if < 0.3.

\textbf{Available Tools} --- get\_function\_source, get\_callers, get\_callees, search\_code, create\_suspicious\_point.

\textbf{SP Format} --- Describe using control flow, not line numbers: ``In function X, when processing Y, the length parameter flows to memcpy without bounds check.''

\textbf{Remember} --- Report first, let experts verify. Better to report 10 SPs with 3 real bugs than to report 2 SPs and miss 1 real bug.
\end{tcolorbox}
\caption{SP Generator Prompt (abridged; full prompt available in supplementary material)}
\label{fig:sp-prompt}
\end{figure*}

\begin{figure*}[t]
\begin{tcolorbox}[colback=orange!5, colframe=orange!40, title=SP Verifier Prompt (Abridged)]
\small\ttfamily
You are a security researcher filtering out obviously wrong suspicious points.

\textbf{Your Role: FILTER, Deep Verify} --- Filter out wrong SPs through in-depth analysis (truly unreachable, wrong sanitizer type). Let uncertain cases PASS to PoC agent for actual testing. PoC failure is cheap; missing a real bug is expensive.

\textbf{Key Principle}: When in doubt, let it through. Only mark FP when you are 100\% certain.

\textbf{Function Pointer Reachability} --- Static analysis may mark functions as ``unreachable'' when they are actually called via function pointers. Check for: struct method dispatch, callback functions, handler patterns. If function pointer pattern found, the function IS reachable.

\textbf{Strict False Positive Rules} --- Only mark FP when:
\begin{itemize}
\item TRULY UNREACHABLE: No direct call AND no function pointer pattern
\item WRONG SANITIZER: Bug type incompatible with sanitizer
\item 100\% CERTAIN protection exists (bounds checks can be wrong!)
\end{itemize}

\textbf{Verification Steps} --- (1) Check static reachability. (2) Check function pointer patterns if static says unreachable. (3) Verify sanitizer compatibility. (4) Analyze source code. (5) Make judgment.

\textbf{PoC Guidance} --- When passing to PoC agent, provide brief guidance: what input to generate, how to reach the vulnerable code.

\textbf{Available Tools} --- get\_function\_source, get\_callers, get\_callees, search\_code, update\_suspicious\_point.
\end{tcolorbox}
\caption{SP Verifier Prompt (abridged; full prompt available in supplementary material)}
\label{fig:spverify-prompt}
\end{figure*}

\begin{figure*}[t]
\begin{tcolorbox}[colback=orange!5, colframe=orange!40, title=PoC Generator Prompt (Abridged)]
\small\ttfamily
You are a security researcher generating POV (Proof of Vulnerability) inputs to trigger a specific vulnerability.

\textbf{Your Task} --- Generate binary input (blob) that triggers the identified vulnerability and causes a sanitizer-detectable crash.

\textbf{Target Configuration}
\begin{itemize}
\item \textbf{Fuzzer}: Defines the INPUT FORMAT your blob must match
\item \textbf{Sanitizer}: Defines what CRASH TYPES can be detected (ASan: buffer overflow, UAF; MSan: uninitialized read; UBSan: integer overflow, null deref)
\end{itemize}

\textbf{Workflow} --- (1) UNDERSTAND: Read vulnerable function, trace data flow from fuzzer input. (2) DESIGN: Plan what bytes trigger the vulnerability. (3) CREATE: Write generator code that produces 3 DIFFERENT blob variants. (4) VERIFY: Test each blob, iterate if no crash.

\textbf{Generator Code Format} --- Your generate(variant) function receives variant number (1, 2, or 3). Return DIFFERENT blobs for each variant.

\textbf{Available Tools} --- get\_function\_source, get\_file\_content, get\_callers, get\_callees, search\_code, create\_pov, verify\_pov, trace\_pov (available after 15 attempts).

\textbf{Tips} --- Read fuzzer source FIRST. Each variant should try a different approach. Start simple, add complexity. If verify fails, analyze output and adjust.

\textbf{Limits} --- Max 40 create\_pov calls. Each attempt generates 3 variants. trace\_pov available after 15 attempts. Stop when crashed=True.
\end{tcolorbox}
\caption{PoC Generator Prompt (abridged; full prompt available in supplementary material)}
\label{fig:poc-prompt}
\end{figure*}

\end{document}